\documentclass[]{aa} 

\usepackage{graphicx}
\usepackage{ulem,color}
\usepackage[dvipsnames]{xcolor}
\usepackage{txfonts}
\begin{document} 

\newcommand{\Ho}{\mbox{$H_0$}}
\newcommand{\ang}{\mbox{{\rm \AA}}}
\newcommand{\abs}[1]{\left| #1 \right|} 
\newcommand{\kms}{\ensuremath{{\rm km\,s^{-1}}}}
\newcommand{\cmsq}{\ensuremath{{\rm cm}^{-2}}}
\newcommand{\ergs}{\ensuremath{{\rm erg\,s^{-1}}}}
\newcommand{\ergsa}{\ensuremath{{\rm erg\,s^{-1}\,{\AA}^{-1}}}}
\newcommand{\ergscm}{\ensuremath{{\rm erg\,s^{-1}\,cm^{-2}}}}
\newcommand{\ergscma}{\ensuremath{{\rm erg\,s^{-1}\,cm^{-2}\,{\AA}^{-1}}}}
\newcommand{\msyr}{\ensuremath{{\rm M_{\rm \odot}\,yr^{-1}}}}
\newcommand{\nhi}{n_{\rm HI}}
\newcommand{\fhi}{\ensuremath{f_{\rm HI}(N,\chi)}}
\newcommand{\refs}{{\bf (refs!)}}
\newcommand{\Av}{\ensuremath{A_V}}
\newcommand{\J}{J0015$+$1842}
\newcommand{\Jlong}{SDSS\,J001514.82$+$184212.34}
\newcommand{\OIII}{\ion{O}{iii}}
\newcommand{\lya}{Ly-$\alpha$}
\newcommand{\CIV}{\ion{C}{iv}}
\newcommand{\Hb}{H-$\beta$}

\newcommand{\iap}{Institut d'Astrophysique de Paris, CNRS-SU, UMR\,7095, 98bis bd Arago, 75014 Paris, France --- \email{noterdaeme@iap.fr}\label{iap}}
\newcommand{\ioffe}{Ioffe Institute, {Polyteknicheskaya 26}, 194021 Saint-Petersburg, Russia \label{ioffe}}
\newcommand{\obspm}{Observatoire de Paris, LERMA, Coll\`ege de France, CNRS, PSL University, Sorbonne University, 75014, Paris, France \label{obspm}}
\newcommand{\iucaa}{Inter-University Centre for Astronomy and Astrophysics, Pune University Campus, Ganeshkhind, Pune 411007, India \label{iucaa}}
\newcommand{\dawn}{Cosmic Dawn Center (DAWN) and Niels Bohr Institute, University of Copenhagen, Jagtvej 128, DK-2200, Copenhagen N, Denmark \label{dawn}}

\newcommand{\PN}[1]{{\color{orange} PN:~ #1}}
\newcommand{\SB}[1]{{\color{olive} SB:~ #1}}
\newcommand{\FC}[1]{{\color{violet} FC:~ #1}}
\newcommand{\old}[2]{{\color[rgb]{0,0,0}\sout{#1}}{\color[rgb]{0.7,0,0.0}{\bf #2}}}
\newcommand{\new}[1]{{\bf #1}}

 \title{
 Remarkably high mass and high velocity dispersion of molecular gas associated with 
 a regular, absorption-selected type-I quasar} 
 \titlerunning{Wide CO(3-2) emission associated with quasar \J}

   \author{
   P. Noterdaeme\inst{\ref{iap}}
   \and
   S. Balashev\inst{\ref{ioffe}}
   \and
   F. Combes\inst{\ref{obspm}}
   \and \\
   N. Gupta\inst{\ref{iucaa}}
   \and
   R. Srianand\inst{\ref{iucaa}}
   \and
   J.-K. Krogager\inst{\ref{iap}}
   \and
   P. Laursen\inst{\ref{dawn}}
   \and
   A. Omont\inst{\ref{iap}}
          }

   \institute{\iap \and \ioffe \and \obspm \and \iucaa \and \dawn}

   \date{\today.}

 \abstract{
We present 3-mm observations of the quasar \J\ at $z=2.63$ with the NOrthern Extended Millimeter Array (NOEMA). Our data reveals molecular gas, traced via a Gaussian CO(3-2) line, with a remarkably large velocity dispersion ($FWHM=1010\pm120$~\kms) {and corresponding to a total molecular mass $M_{\rm H_2} \approx (3.4\text{--}17) \times 10^{10} M_{\odot}$, depending on the adopted 
CO-to-H$_2$ conversion factor $\alpha_{\rm CO}=(0.8\text{--}4.0)~{M_{\odot}}\,(\kms\,{\rm pc}^2)^{-1}$.}
Assuming the 3-mm continuum emission is thermal, we derive a dust mass of {the order of} $M_{\rm dust} \sim 5 \times 10^8 M_{\rm \odot}$. 
\J\ is located in the molecular gas-rich region in the IR vs CO line luminosity diagram, in-between the main locus of main-sequence and sub-millimetre galaxies and that of most other active galactic nuclei (AGNs) targeted so far for CO measurements.
While the large velocity dispersion of the CO line suggests a {merging system},  
\J\ is observed to be a regular, only very moderately dust-reddened ($A_V \sim 0.3\text{--}0.4$) type-I quasar from its UV-optical spectrum, from which we infer a mass of the super-massive black hole be around $M_{\rm BH} \approx 6\times 10^8 M_{\rm \odot}$.\\
We suggest that \J\ is observed at a galaxy evolutionary stage where a massive merger has brought  
significant amounts of gas towards an actively accreting super-massive black hole (quasar). 
{While the host still contains a large amount of dust and molecular gas with high velocity dispersion, 
the quasar has already cleared the way \textit{towards the observer}, likely through powerful outflows as recently revealed by optical observations of the same object.}
\\ {High angular resolution observations of this and similar systems} should help determining better the respective importance of evolution and orientation in the appearance of quasars and their host galaxies and have the potential to investigate early feedback and star-formation processes in galaxies in their quasar phases.
 }

\maketitle

\section{Introduction}

There is accumulating evidence for a strong link between the evolution of massive galaxies and the super-massive black hole (SMBH) that they generally host in their centre \citep[e.g.][]{Heckman2004}. When matter is accreted onto the disc surrounding the SMBH, enormous amounts of energy can be released through radiation (or relativistic jets), triggering the so-called galactic nuclear activity.  Major mergers have been proposed as an efficient mechanism to make the matter lose most of its angular momentum and fall down to the inner regions of the galaxy in a relatively short time scale (e.g. \citealt{Silk1998,Volonteri2003,Springel2005}, but see \citealt{Miki2021}). At the same time, merging systems are known to induce intense star-formation activity through compression of the gas. In fact, starburst galaxies and luminous unobscured active nuclei (quasars) are possibly the same systems observed at different stages of the galaxy-SMBH co-evolution \citep{Hopkins2008}. In the rapid SMBH growth phase, huge amounts of dust and molecules are brought to the galaxy centre, and the SMBH activity remains heavily obscured. Bright, unobscured quasars would then correspond to 
a later phase, when the SMBH have almost fully assembled and are radiating close to their Eddington limit. The energy released could then be sufficient 
to significantly clear dust and gas from the entire galaxy through powerful winds. Such a feedback mechanism could be responsible for quenching star formation 
in the host \citep[e.g.][]{Zubovas2012,Pontzen2017,Terrazas2020}, although the observed large-scale outflows may not necessarily arise from propagation of energy from the accretion disc to the interstellar medium \citep{Fabian2012,Veilleux2017}, i.e. powered by AGN activity, but could also result from 
the intense star-formation activity or tidal ejection during the merging phase \citep[e.g.][]{Puglisi2021}.

Since cold molecular gas is expected to support both the star-formation and the growth of the SMBHs, large efforts have been devoted to study the properties of this phase {\citep[e.g.][among many other works]{Omont1996,Barvainis1997,Lewis2002,Bertoldi2003,Riechers2006,Weiss2007,Wang2010,Salome2012}}. 
In particular, many observing campaigns aimed at detecting the CO emission lines and constrain the molecular reservoirs of galaxies to investigate the evolutionary sequence between starburst (SB) galaxies and quasars, {as well as AGN feedback (either positive or negative) on star-formation in the host \citep[e.g.][]{Weiss2012,Nesvadba2020}.} 
The studies have mostly focused on the bright end of the infra-red luminosity distribution and 
showed that luminous quasars generally have low ratio of molecular gas masses to star-formation rates (SFR) \citep[e.g.][]{Bischetti2021}, while starburst galaxies have much larger molecular masses to SFR ratios, supporting the evolutionary paradigm. 
Several studies have also searched for CO emission in obscured quasars, i.e. possibly at the short-lived intermediate stage between the SB and optically-bright quasar phase (e.g. {\citealt{Polletta2011}}, \citealt{Brusa2015}), but differences with luminous quasars seem to appear only for Compton-thick cases, more likely associated to the initial steps of the blow-out phase \citep{Perna2018}.

On the other hand, according to the AGN unification scheme \citep{Antonucci1993}, a given AGN appears obscured or not depending 
on its orientation with respect to the observer, i.e. whether the line of sight crosses large amounts of dust in the circum-nuclear region or not. It is therefore not absolutely clear how much the appearance of a system depends on its evolutionary 
stage and its orientation. It is also possible that the disc of the host galaxy also contributes significantly to obscuring 
the nuclear region \citep[e.g.][]{Gkini2021}. 

Recently, \citet{Noterdaeme2019} uncovered a population of unobscured quasars in the Sloan Digital Sky Survey \citep[SDSS][]{York2000} featuring strong H$_2$ absorption lines at the quasar redshift 
in its optical spectrum. 
Detailed spectroscopic investigations of one of them, \Jlong\ (hereafter \J) with X-shooter on the Very Large Telescope, suggested that the absorption system belongs to a galactic-scale multi-phase outflow, which is also revealed by spatially resolved [\OIII] and \lya\ emission \citep{Noterdaeme2020}. In this Letter, we present the detection of strong and remarkably broad CO(3-2) emission with the NOrthern Extended Millimeter Array (NOEMA) in this quasar. We discuss these exceptional characteristics for an otherwise apparently regular quasar and suggest that the orientation of this system allows us to observe early AGN feedback {during or} after a merger phase.

\section{Observations}

Observations were carried out with  the NOrthern Extended Millimeter Array (NOEMA) and the PolyFIX correlator in the 3-mm band 
in May and June, 2020. The CO(3-2) line at $z=2.631$ is redshifted at the 95.234 GHz frequency. We observed in D-configuration, with 10 antennas on 23 and 29 May, and 9 antennas on June 3. On May 23$^{\rm rd}$, the weather conditions were good and stable and we obtained 2.4~h of integration. On May 29$^{\rm th}$, most of the data was flagged because of poor weather, leaving 0.8~h of integration. On June 3$^{\rm rd}$, the weather was fine, providing us 4.9~h of integration time. Calibration was done using 5 sources, 3C345, 3C454.3, 0007+171, 2010+723 and MWC349. The absolute flux calibration is accurate at
the 10\% level. The data were calibrated with the CLIC package and mapped with the MAPPING package in the GILDAS software\footnote{\url{http://www.iram.fr/IRAMFR/GILDAS/}}.  Using CLARK cleaning in natural weighting, the compact D configuration provided a beam of
 3\farcs9$\times$3\farcs7, with a PA of 69$^\circ$ 
 for the CO(3-2) line and the upper sideband continuum and 
 5\farcs1$\times$4\farcs5, with a PA of 125$^\circ$ for the lower sideband continuum. The continuum was also computed 
 with the wider 7.7 GHz upper side band, with a beam of 3\farcs9$\times$3\farcs4, with a PA of $-53^\circ$.

 The quasar was observed
in dual polarisation mode in 4 basebands, with 3.9~GHz total bandwidth per baseband, distributed in lower and upper sidebands distant by 15.5 GHz. The CO(3-2) line was observed in the upper side-band, and no other line was detected in the rest of the basebands (upper and lower), which were used to estimate the continuum level. The velocity resolution was initially 2~MHz $\sim$6.29~\kms. The spectra were then smoothed to 38.13~MHz (120~\kms) to build channel maps.
The final cube is 128$\times$128 pixels with 0\farcs785 per pixel in the plane
of the sky, and has 80 channels of 120~\kms width.
The noise level is 175 $\mu$Jy/beam in 120~\kms\ channels for the line and {14 $\mu$Jy/beam} for the continuum.

\section{Results}

\begin{figure*}
    \centering
    \addtolength{\tabcolsep}{-4pt}
    \begin{tabular}{cccc}
    \includegraphics[trim=3.3cm 7.5cm 16.95cm 5cm,clip,height=0.24\hsize]{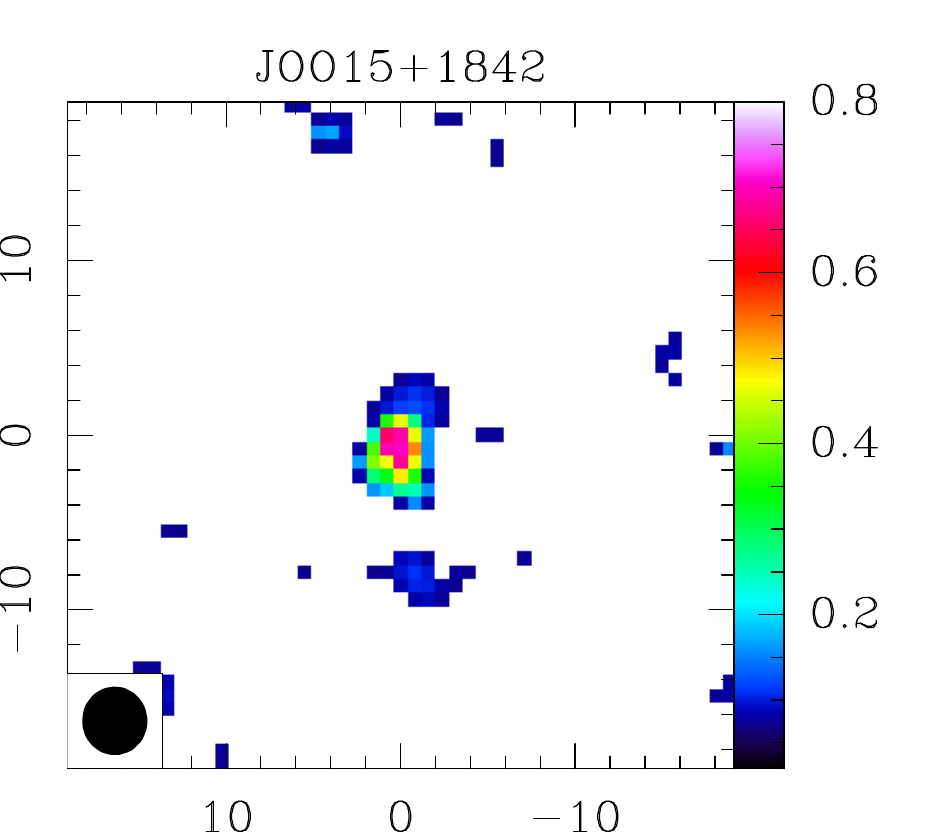}&
    \includegraphics[trim=3.95cm 7.5cm 16.95cm 5cm,clip,height=0.24\hsize]{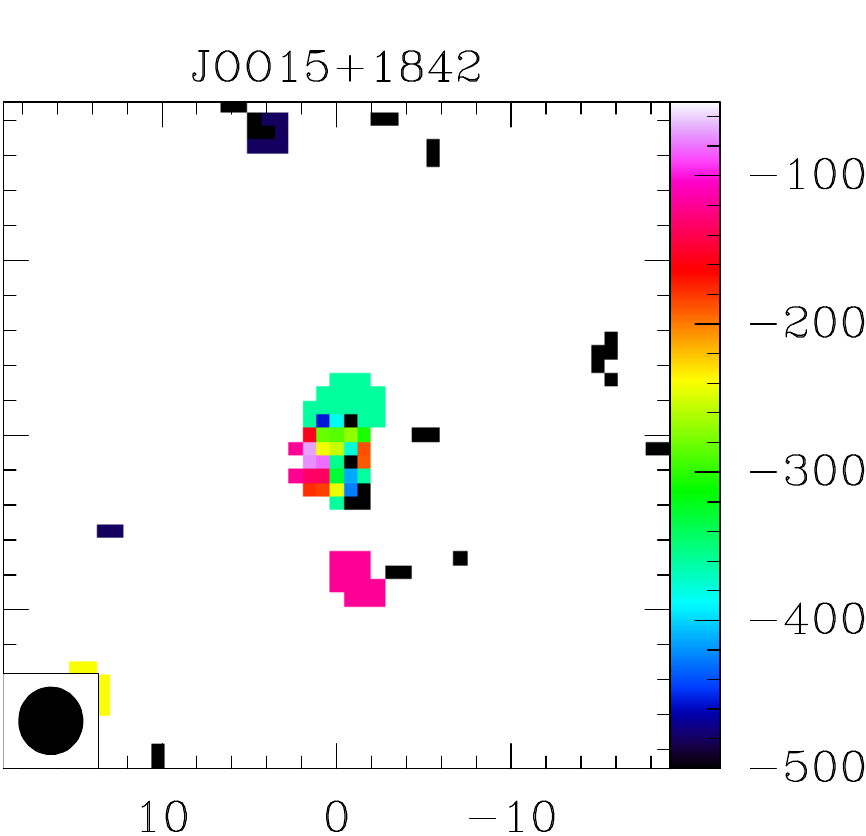}&
    \includegraphics[trim=3.95cm 7.5cm 16.95cm 5cm,clip,height=0.24\hsize]{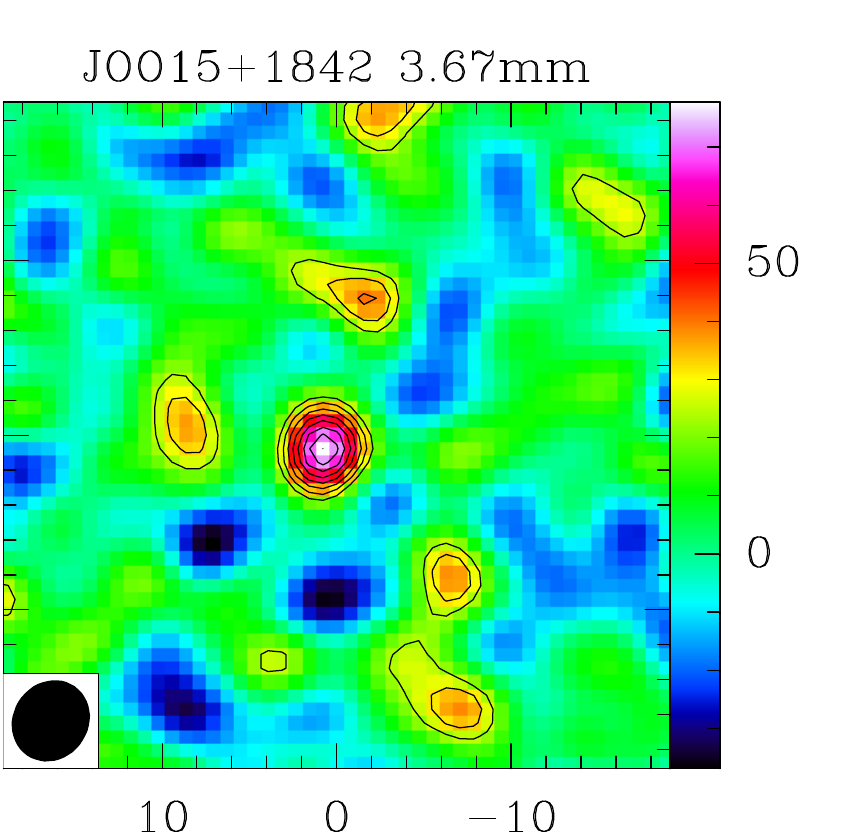}&
    \includegraphics[trim=3.95cm 7.5cm 16.95cm 5cm,clip,height=0.24\hsize]{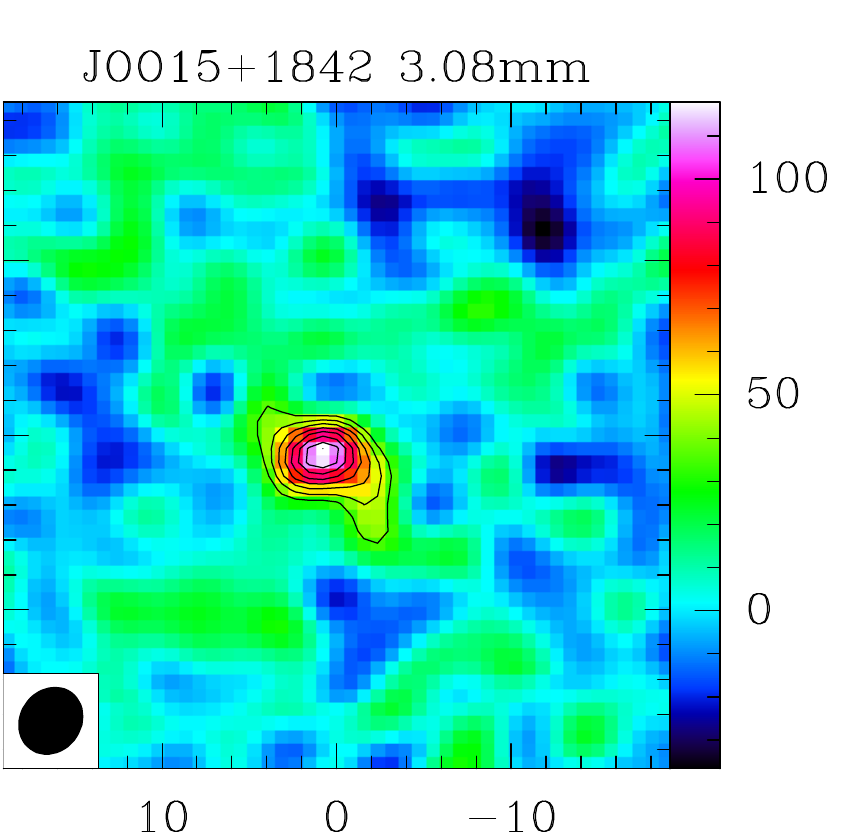}\\
    \end{tabular}
    \caption{From left to right: Integrated CO(3-2) map, with colour scale in Jy/beam$\times$km/s; 
    CO(3-2) velocity map, with colour scale in \kms; 
   continuum maps in the lower (3.67\,mm) and the upper (3.08\,mm) sidebands with colour scale in $\mu$Jy/beam. {Contours are from 30 to 100\% by 10\% of the maximum, which is 78$\mu$Jy/beam and 118$\mu$Jy/beam respectively.}
All axes are in offset arcseconds from the central position RA=00:15:14.81	Dec=18:42:12.30.}
    \label{f:maps}
        \addtolength{\tabcolsep}{+6pt}
\end{figure*}

Fig.~\ref{f:maps} presents the CO(3-2) integrated map, the velocity dispersion map, and the continuum maps in the two sidebands. The moments of the line cube have been taken above a threshold of 3$\sigma$ in 120 km/s channels, corresponding to an integrated level of 0.063 Jy/beam km/s. 

\subsection{CO(3-2) emission}

The CO(3-2) emission is clearly detected in the integrated map, but the source remains unresolved, given the large beam size of 4\arcsec, i.e. corresponding to 32~kpc at the quasar's redshift. 
{The one moment map of Fig.~\ref{f:maps} presents a possible East-West velocity gradient, when the outflowing gas has likely a North-South direction projected on the sky \citep{Noterdaeme2020}.}

\begin{figure}
    \centering
    \includegraphics[width=0.95\hsize]{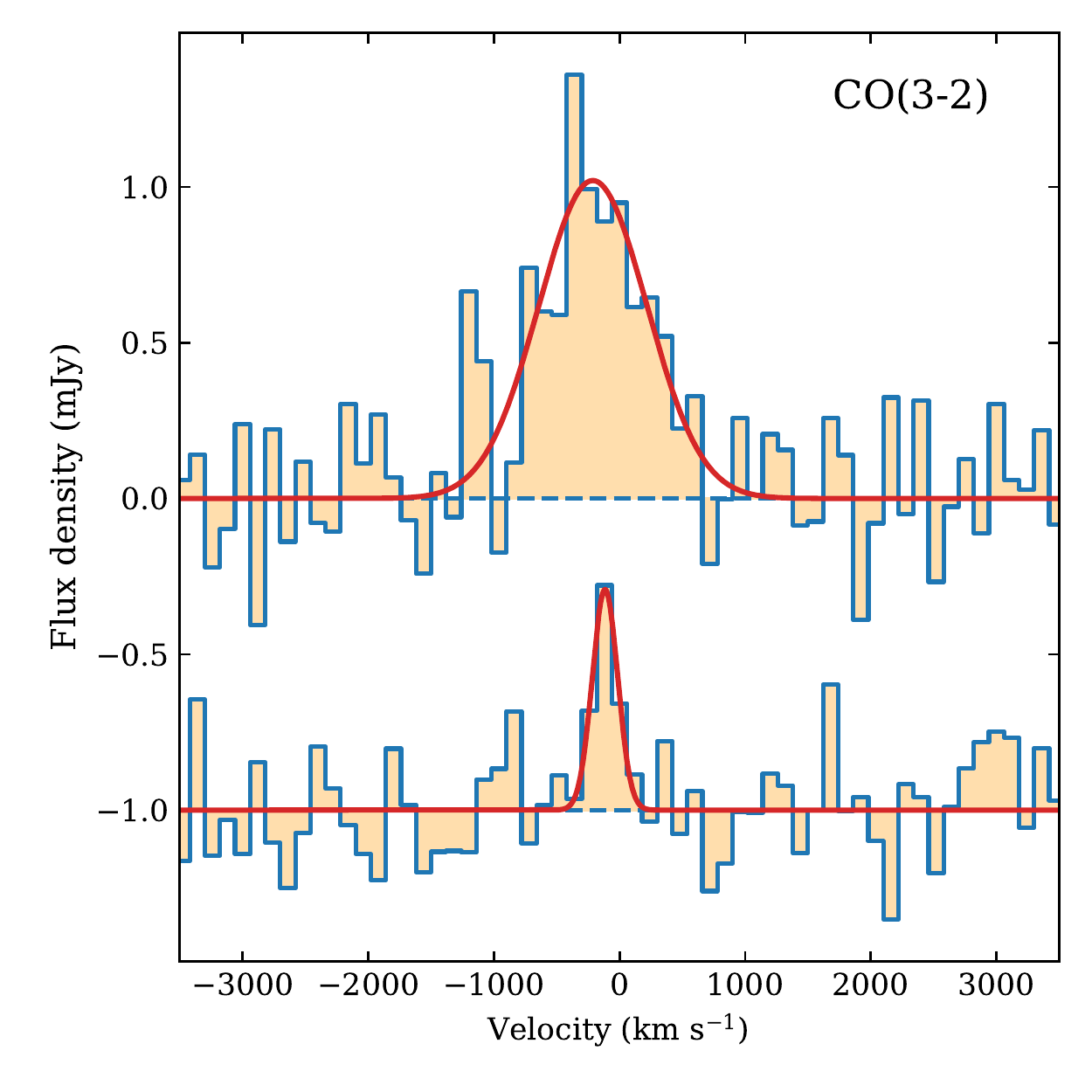}
    \caption{The CO(3-2) spectrum extracted using 120~\kms~bins {within a circle of 4'' diameter, i.e. an} aperture as large as the synthesised beam of our NOEMA observations {(top)}. {The spectrum extracted at pixel (-0.78, -7.85\arcsec) is also shown, shifted by -1 mJy for convenience.}
    The zero of the velocity scale corresponds to the systemic redshift $z=2.631$ inferred from the narrow 
    [O\,{\sc iii}] emission \citep{Noterdaeme2020}.}
    \label{f:spec}
\end{figure}

In Fig.~\ref{f:spec}, we present the 3mm spectrum extracted at the position of the quasar. We fitted the CO(3-2) 
line with 
a Gaussian function, yielding a total integrated flux $F_{\rm CO(3-2)} = 1.1 \pm 0.2$~Jy\,\kms, 
$FWHM = 1010 \pm 120$~\kms\ and a central velocity $v=-210 \pm 50$\,\kms\ with respect to {the systemic redshift}\footnote{{$z=2.631$, inferred with $\sim 100$~\kms\ uncertainty from the narrow [\OIII] emission.}}. {Such shift, if real, may be due to blending of CO lines in a complex system.} 
{We converted the flux to luminosity following \citet{Solomon2005} and obtained $L'_{\rm CO(3-2)} \approx 4.1\times 10^{10}$~K\,km\,s$^{-1}$\,pc$^2$.}
{To estimate the mass of molecular gas from the CO line luminosity, we 
assumed a typical CO spectral line energy distribution for quasars and took the CO(3-2)/CO(1-0) intensity ratio $r_{31}=0.97$ from \citet{Carilli2013}, consistent with observations of AGN host galaxies at $z>2$ \citep{Sharon2016}. 
The CO-to-H$_2$ conversion factor depends on the average conditions in the molecular gas which we do not know.  We hence obtain a conservative range $M_{\rm H_2} =\alpha_{\rm CO}\,L'_{\rm CO(1-0)}\approx (3.4-17)\times 10^{10} M_{\rm \odot}$ assuming $\alpha_{\rm CO} = 0.8~{M_{\odot}}\,(\kms\,{\rm pc}^2)^{-1}$ (as generally adopted for quasars \citealt[e.g.][]{Walter2003, Wang2010,Bolatto2013}) and $\alpha_{\rm CO} = 4~{M_{\odot}}\,(\kms\,{\rm pc}^2)^{-1}$ (standard value).}

{Finally, we also note a possible structure about 8\arcsec ($\sim$60~kpc at $z=2.631$) southwards of the quasar in both the zero and one moment maps. This could be due to a gas-rich companion galaxy not seen in the optical. The corresponding extracted spectrum has $F_{\rm CO(3-2)} = 0.18 \pm 0.06$~Jy\,\kms, $FWHM = 240 \pm 60$~\kms\ and $v=-115 \pm 30$\,\kms. This corresponds to a molecular mass of 2.8$\times 10^{10} M_{\rm \odot}$, assuming the same $r_{31}$ as above and the standard $\alpha_{\rm CO}$. Deeper observations with higher spatial resolution are necessary to better unveil the quasar and its environment.}

\subsection{Continuum 3\,mm emission and FIR luminosity}

We detect the 3~mm continuum emission at the position of the quasar 
on the map with a flux density, {determined by fitting uv visibilities, of 
 65$\pm$14\,$\mu$Jy and 104$\pm$14\,$\mu$Jy} for the lower (81.76\,GHz or 3.67\,mm) and upper (97.24\,GHz or 3.08\,mm) sidebands, respectively.
The flux density is varying with frequency as a power-law with slope of $\nu^\alpha$ with 
$\alpha=2.7$, i.e. consistent with dust emission, with a dust opacity varying as $\nu^\beta$, with $\beta = 0.7$ albeit with a large ($\pm 1.2$) {statistical} uncertainty\footnote{{This clearly dominates over systematics since the relative flux calibration of the basebands (USB,LSB) of NOEMA is very good \citep{Neri2020}.}}. 
Following \citet[][their Eq.~2]{Carniani2017}{, but assuming a dust temperature in the range 40-80~K and $\beta$ within the range obtained above, we infer a dust mass of $M_{\rm dust} \approx 5\times10^8 M_{\rm \odot}$, within a factor of two, assuming thermal emission only.}

\begin{figure}
    \centering
   \includegraphics[width=\hsize]{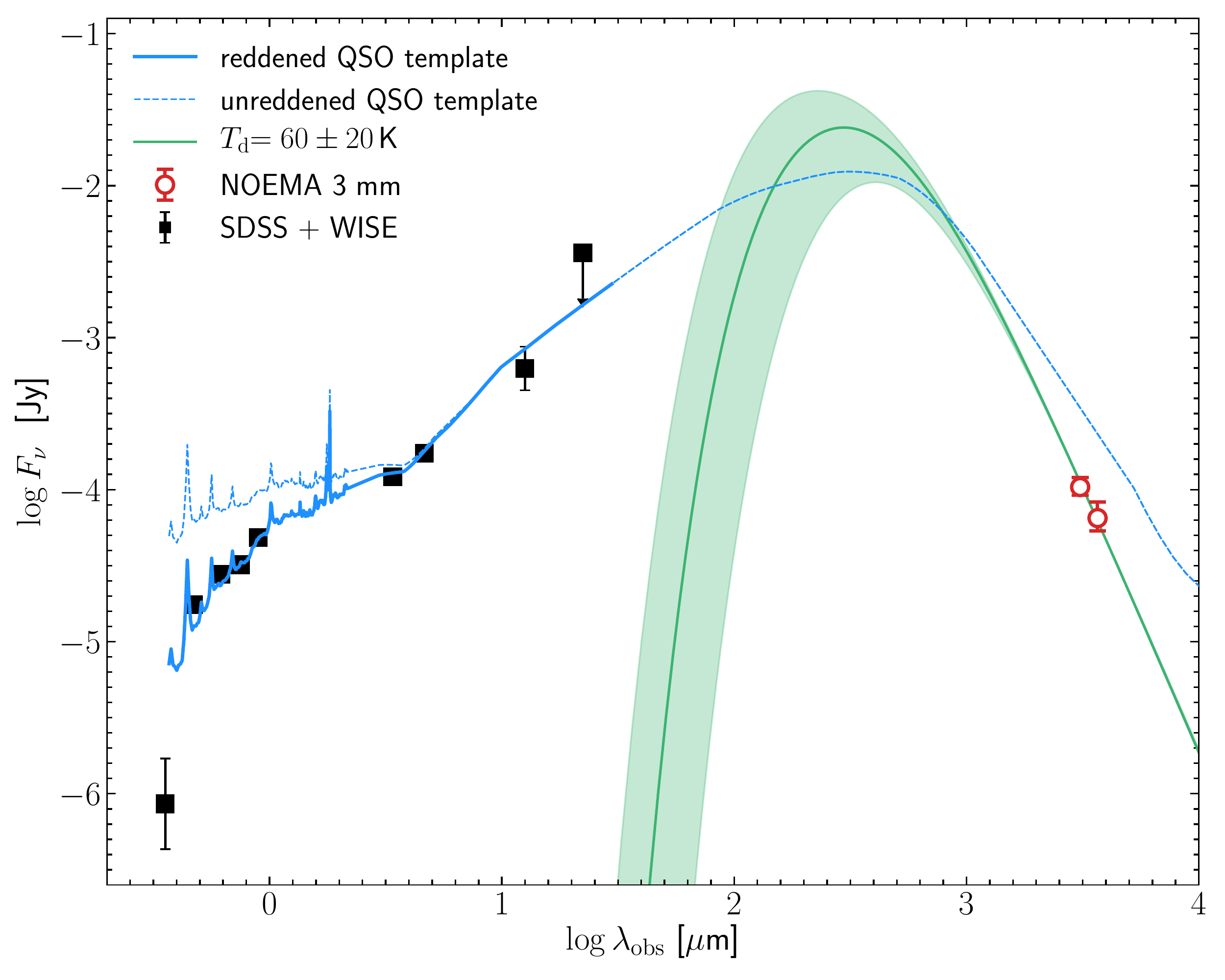}
    \caption{Fit to the spectral energy distribution of the quasar \J\ using photometric points from SDSS and WISE (black squares). The solid {(resp. dotted)} blue line shows the reddened {(resp. unreddened)} quasar type I template from \citet{Polletta2007}. The NOEMA continuum measurements at 3.08 and 3.67~mm are represented by the red circles with error bars, and the associated constraint on the cold dust emission component assuming $T_d=60\pm20$~K is shown in green. 
    }
    \label{f:sed}
\end{figure}

We constrained the AGN spectral energy distribution (SED) from the fluxes measured in SDSS and Wide-field Infrared Survey Explorer \citep[WISE,][]{Wright2010} filters\footnote{The $u$-band is not considered at it is affected by \lya\ forest and H$_2$ absorption. W4 provides an upper limit. The quasar presents only $\sim$0.25 V-mag variability according to the Catalina Sky Survey \citep{Drake2009}, consistent with the long-term variability of other quasars \citep{Hook1994}.}, using the template by \citet{Polletta2007}, 
only moderately reddened by dust {($A_V\sim 0.3$, consistent with $A_V=0.4\pm0.1$ derived from the X-shooter spectrum and template, \citealt{Noterdaeme2020})}, see Fig.~\ref{f:sed}.  
{Based on this template, the intrinsic bolometric luminosity is found to be $\log L_{\rm bol} / L_{\odot} \simeq 13.4$.} 
{We note that the relative contribution of the host galaxy is probably different than predicted by this template which over-predicts the 3\,mm continuum emission.}
{We fitted this emission using a modified black-body emission law \citep[see e.g. Eq.~2 from][]{Rangwala2011}, as expected for reprocessed cold dust emission}.
{Using standard values for power law slope ($\beta=1.6$, consistent with our constraint) and normalisation point of the frequency dependence of the effective optical depth ($\nu_0 = 1.5$\,THz), and assuming a} dust temperature 
$T_{\rm d}=60$\,K, 
we get a total IR luminosity associated with the host 
$\log L_{\rm IR}/L_{\odot} \sim 12.7$. 
We caution however that this estimate is very uncertain in the absence of measurements
at $100\text{--}1000\,\mu$m to constrain $T_{\rm d}$ and $\beta$. 
{Assuming $T_{\rm d}$ in the range 40-80~K results in about one order of magnitude uncertainty of $\log L_{\rm IR}/L_{\odot}=12.2-13.1$.} 
{We also note that if a merging/companion galaxy contributes to this IR luminosity \citep{Bischetti2021}, then the value could be considered as an upper-limit to the IR luminosity of the quasar host alone.}

\section{Discussion and conclusions}
In order to maximise the chances of detection, the vast majority of $z>1$ AGNs observed so far in CO lines have been selected for being among the brightest objects, in particular at millimetre or infra-red wavelengths (e.g. \citealt{Coppin2008,Wang2010,Simpson2012}; {\citealt{Wang2013}}; \citealt{Feruglio2014,Wang2016};{ \citealt{Fan2019,Banerji2021}}). 
For example, WISE has played an important role in selecting these hyperluminous objects, in combination \citep[e.g.][]{Bischetti2017} or not \citep[][]{Fan2018} with photometry from the SDSS. Other works focused on possibly less luminous objects, but still preferentially dust-obscured \citep[e.g.][]{Polletta2011,Banerji2017,Kakkad2017,Perna2018}, radio-loud \citep[e.g.][]{Willott2007} or on the highest redshifts detected in [\ion{C}{II}]$\lambda$158$\mu$m \citep[e.g.][]{Venemans2017}.  
In turn, \J\ is not particularly bright, with e.g. $i$-band magnitude in only the second brightest quartile of SDSS quasars at the same redshift, and only moderately reddened {($A_V \sim 0.3-0.4$)}. It has also $\sim$3 mag fainter WISE magnitudes than those of WISSH quasars \citep{Bischetti2017}. 

\begin{figure}
    \centering
    \includegraphics[width=\hsize]{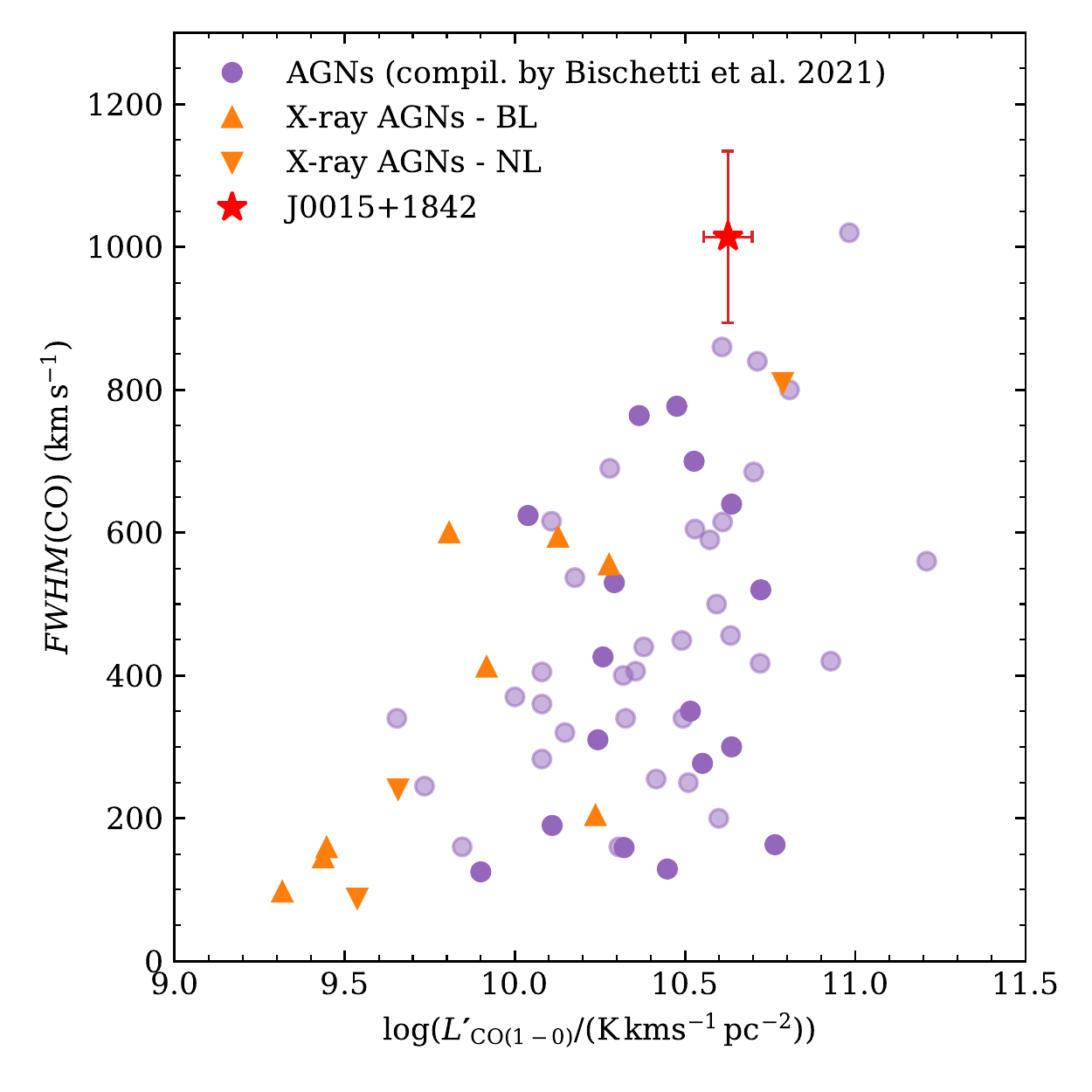}
    \caption{Full-width at half maximum versus luminosity of the CO emission line in \J\ (red star) compared to that in $z>1$ active galactic nuclei (compiled by \citet{Bischetti2021}, with X-ray selected broad-line (BL) and narrow-line (NL) AGNs from \citet{Circosta2021} shown by orange triangles). Measurements using CO(3-2) are shown with darker symbols than those from other CO lines. 
    }
    
    \label{f:FWHM}
\end{figure}

\begin{figure}
    \centering
    \includegraphics[width=\hsize]{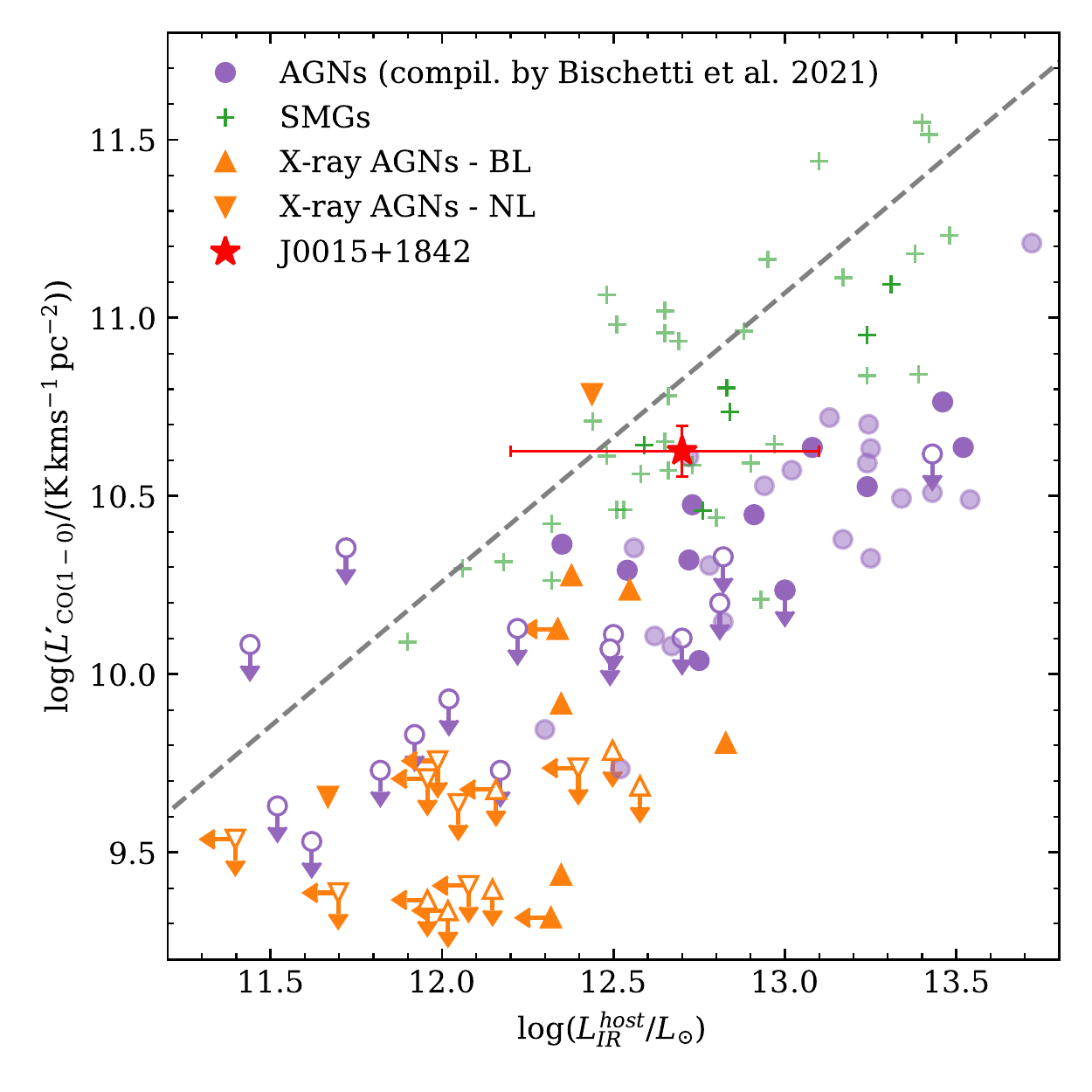}
    \caption{CO line luminosity against the infra-red luminosity of the host integrated in the $8\text{--}1000\,\mu$m range. 
      Unfilled symbols corresponding to CO upper limits in case of non-detection and green crosses to sub-millimetre galaxies. Other symbols are as per Fig.~\ref{f:FWHM}. The dashed line corresponds to the relation for main-sequence galaxies \citep[][]{Sargent2014}. }
    \label{f:COIR}
\end{figure}

In Fig.~\ref{f:FWHM}, we compare the width and luminosity of the detected CO line with those seen in other $z>1$ AGNs from the literature. 
We use the compilation of unlensed AGNs by \citet{Bischetti2021} and retrieved the line widths from the original detection publications. In a few cases where several {spectrally resolved components}
were reported, we considered conservatively the widest one. 
To avoid systematics related to the assumption of CO-to-H$_2$ conversion factor, we compare directly the CO line luminosity instead of the H$_2$ mass. This compilation also includes the sample by \citet{Circosta2021} who 
recently performed a X-ray selection of both type I and type II AGNs 
with a range of bolometric luminosities, independently on their mm or IR emission. This more homogeneous sample also presents the advantage of 
being observed all in the CO(3-2) line, so that direct comparison can be made with \J\, regardless of the assumption of the excitation correction. We therefore distinguish this sample from other AGNs in Figs.~\ref{f:FWHM} and \ref{f:COIR}. 

Both the luminosity and the line width of \J\ are remarkably high, and the width measured here, $FWHM = 1010\pm120$~\kms\ is well above the vast majority of other measurements in the compiled sample, only equated by {the powerful obscured quasar SWIRE\,J022513-043419 that has $FWHM\approx1020\pm110$~\kms\ \citep{Polletta2011}. Like \J, the latter object also presents extended [\OIII] emission that may trace outflowing gas \citep{Nesvadba2011}.
}

We compare the CO content of \J\ with other AGNs as a function of the IR luminosity 
in Fig.~\ref{f:COIR}, using the same compilation as previously, but restricted to systems for which \citet{Bischetti2021} consider the IR measurement to be reasonably reliable (their section 6.2). We also add sub-millimetre galaxies (SMGs) to the comparison.
The IR luminosity, integrated over the range $8\text{--}1000\,\mu$m, is generally considered as a proxy of dust emission related to star formation in the host, once the AGN contribution has been removed. The IR-to-CO luminosity ratio is then a widely used proxy for star-formation efficiency. 
{Remarkably, the high CO-line luminosity of \J\ is similar to that of the most luminous AGNs in IR.} 
{However, with its likely lower cold dust emission, \J\ is located closer to the region 
populated by MS galaxies and SMGs, and with molecular content well above that of most X-ray selected AGNs.} The only AGN in the sample of \citet{Circosta2021} with similar CO luminosity is cid\_1253 (COSMOS J100130.56+021842.6) with $\log(L'{\rm _{CO(3-2)}/(K\,\kms\,pc^{-2}})) = 10.80\pm0.04$ and $FWHM({\rm CO})=810\pm93~\kms$. This object is a {merger, hosting a} narrow-line (type II) AGN, while \J\ is a regular quasar with broad emission lines (type I).

According to the evolutionary scheme, major mergers of gas-rich galaxies trigger starburst activity and bring a significant amount of matter towards the nuclear regions, initiating the build up of the supermassive black hole. These objects would mostly appear first as obscured AGNs, with star-formation efficiencies much higher than in main sequence galaxies, then transition to optically bright systems with low gas fractions. Indeed, the presence of powerful multiphase galactic-scale outflows show that AGNs are capable of removing large amounts of gas from their host galaxies. As recently discussed by \citet{Perna2018}, neither the evolutionary sequence nor orientation effects with respect to obscuring medium alone are able to explain the observed SFEs in obscured and unobscured AGNs. 

\J\ is bringing an important piece to this puzzle. The exceptionally broad Gaussian profile of the CO line in \J\ suggests {a recent or ongoing} merger, with a very significant amount of molecular gas and dust. {However,  the line of sight to the active nucleus is only very moderately reddened and the presence of broad emission lines corresponds to a regular type-I quasar} powered by a highly accreting super-massive black hole 
with Eddington ratio\footnote{$L_{\rm Edd} = 1.4\times10^{38}(M_{\rm BH}/M_{\odot})$ and $\log M_{\rm BH}/M_{\rm \odot}\sim 8.6\text{--}8.9$ is obtained from the width of the \CIV\ (or \Hb) line and the rest-frame continuum 
luminosity at 1350~{\AA} (\Hb\ luminosity) using the calibration by \citet{Vestergaard2006}.} $\lambda_{\rm Edd} = L_{\rm bol}/L_{\rm Edd} \gtrsim 1$.

From the analysis of ionised emission lines together with absorption from H$_2$, \citet{Noterdaeme2020} suggest that a multi-phase outflow is observed oriented almost towards the observer. {This could provide a natural explanation to the low extinction along the line of sight.} 
\J\ supports then a picture in which feedback processes can start early in the evolutionary sequence, 
{with outflows} clearing the view towards the nuclear region {at least in some directions}, 
while a large amount of molecular gas is still available in the host. {In addition, although requiring confirmation,} the star-formation efficiency may not have yet reached a level as high as seen in other, possibly more evolved, quasars. 
Measuring the flux density at $\sim$1~mm {would help measuring better} 
the IR luminosity and host SFR of \J\ and similar sources. 

If, as suggested by \citet{Noterdaeme2020}, the presence of proximate H$_2$ absorbers and 
leaking \lya\ emission provide an efficient way to identify multi-phase outflows in regular quasars, then observations of CO emission in a sample of them could bring further clues to constrain the relative importance of orientation and evolutionary sequence in the appearance of quasars. {Deep observations at high angular resolution should allow one} to confirm or not the derived configurations {and} enable detailed investigation of early feedback mechanisms in more regular quasars than those usually targeted for millimetre studies.

\begin{acknowledgements}
{We thank the referee for insightful comments that helped improving the presentation 
of our results.} 
We warmly thank Manuela Bischetti for kindly proving us with her compilation of CO and IR measurements for AGNs and SMGs and Pierre Cox for advises about NOEMA observations.
This work is based on observations carried out under project number S20BZ (PI Noterdaeme) with 
the IRAM NOEMA Interferometer. IRAM is supported by INSU/CNRS (France), MPG (Germany) and IGN (Spain). We are grateful to the IRAM staff for help with the data processing. 
The research leading to these results has received support from the French \textit{Agence Nationale de la Recherche} under 
ANR grant 17-CE31-0011-01/project ``HIH2'' and from the French-Russian Collaborative Programme 1845. SB was supported by RSF grant 18-12-00301.
The Cosmic Dawn Center is funded by the Danish National Research Foundation under grant No.~140.

\end{acknowledgements}

\bibliographystyle{aa}
\bibliography{mybib.bib}
\end{document}